\documentclass[prl,superscriptaddress,twocolumn]{revtex4-2}
\usepackage{graphicx,epsfig,amsmath,amsfonts,fancyhdr,amssymb,xcolor,braket,subcaption,ragged2e,comment}
\usepackage[colorlinks=true]{hyperref}
\usepackage{pdfpages}

\DeclareCaptionJustification{justified}{\justifying}
\newcommand{\qprop}{Q\textsc{prop}}

\newcommand{\figref}[1]{\figurename~\ref{#1}}

\renewcommand{\eqref}[1]{Eq.~(\ref{#1})}
\newcommand{\cpeq}{=}
\newcommand{\changes}[1]{#1}
\newcommand{\changesTwo}[1]{#1}

\DeclareMathOperator{\sign}{sign}

\makeatletter
\AtBeginDocument{\let\LS@rot\@undefined}
\makeatother
\begin{document}

\title{Ultrafast imaging of molecular chirality with photoelectron vortices}
\author{Xavier Barcons Planas} 
\affiliation{ICFO-Institut de Ciencies Fotoniques, The Barcelona Institute of Science and Technology, 08860 Castelldefels (Barcelona), Spain}
\author{Andrés Ordóñez}
\affiliation{ICFO-Institut de Ciencies Fotoniques, The Barcelona Institute of Science and Technology, 08860 Castelldefels (Barcelona), Spain}
\author{Maciej Lewenstein}
\affiliation{ICFO-Institut de Ciencies Fotoniques, The Barcelona Institute of Science and Technology, 08860 Castelldefels (Barcelona), Spain}
\affiliation{ICREA, Pg. Lluís Companys 23, 08010 Barcelona, Spain}
\author{Andrew S. Maxwell}
\affiliation{ICFO-Institut de Ciencies Fotoniques, The Barcelona Institute of Science and Technology, 08860 Castelldefels (Barcelona), Spain}
\affiliation{Department of Physics and Astronomy, Aarhus University, DK-8000 Aarhus C, Denmark}

\date{\today}

\begin{abstract}
Ultrafast imaging of molecular chirality is a key step towards the dream of imaging and interpreting electronic dynamics in complex and biologically relevant molecules. Here, we propose a new ultrafast chiral phenomenon exploiting recent advances in electron optics allowing access to the orbital angular momentum of free electrons. We show that \changes{strong-field} ionization of a chiral target with a few-cycle linearly polarized 800 nm laser pulse yields photoelectron vortices, whose chirality reveals that 
of the target, and we discuss the mechanism underlying this phenomenon. \changes{Our work opens new perspectives in recollision-based chiral imaging.}

\end{abstract}

\maketitle

A major goal motivating research in ultrafast science \cite{corkum2007attosecond, krausz2009attosecond} is to image and understand electron dynamics in complex systems, such as biologically
relevant molecules \cite{calegari_ultrafast_2014}. However, except for the simplest molecules, interpretation of ultrafast measurements in terms
of molecular structure remains extremely challenging \cite{leone_what_2014, xu_time-resolved_2016}. 
Since chirality is both a common and a key structural property of biological molecules \cite{bonner1995chirality}, 
the ultrafast imaging of molecular chirality has become a natural objective of ultrafast science during the last decade \cite{cireasa_probing_2015, beaulieu_attosecond-resolved_2017, baykusheva_chiral_2018, rozen_controlling_2019, ayuso_synthetic_2019}. The current progress in this direction \cite{ayuso_ultrafast_2022} has invariably
relied on the \emph{Archimedes
screw} principle: a chiral structure, here the molecular potential, converts the electron rotation induced by the electric field into
a linear electron current perpendicular to the rotation plane.

\begin{figure*}
    \centering
    \includegraphics[width=\textwidth]{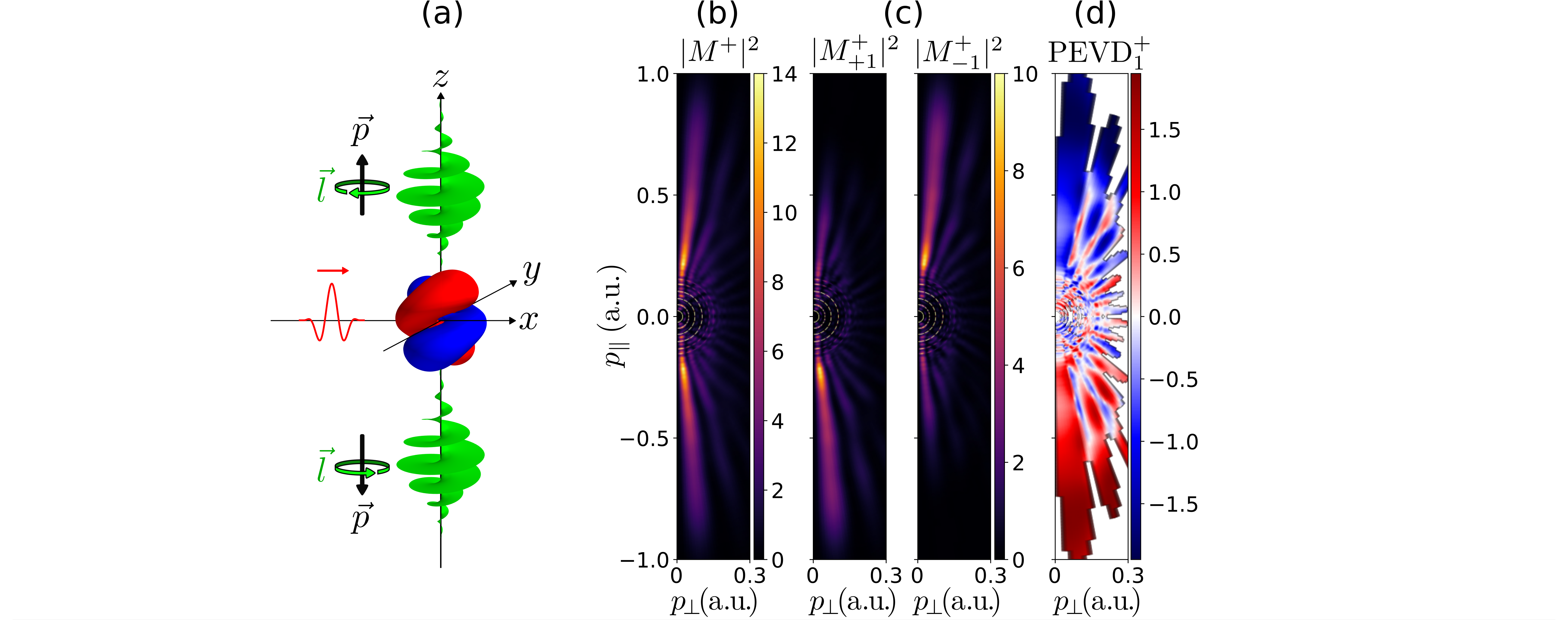}
    \caption{(a) Photoionization of a chiral target with linearly polarized light produces photoelectron vortices with a helicity that encodes the handedness of the target, thus carrying opposite OAM in opposite propagation directions. 
    The target orbital at the origin shows isosurfaces $\chi^+(\vec{r})=\pm10^{-3}$ a.u. [\eqref{eq:rhostate}]. (b) Standard photoelectron spectrum [\eqref{eq:M_planewave}] resulting from strong-field ionization of a perfectly aligned chiral target $\ket{\chi^+}$ [\eqref{eq:rhostate}] as a function of the momentum perpendicular and parallel to the laser field. (c) 
    OAM-resolved photoelectron spectra [\eqref{eq:M_vortex}] for OAM $l_v \cpeq +1$ (left) and $l_v \cpeq -1$ (right). (d) Difference of the OAM-resolved spectra normalized by their average 
    [\eqref{eq:PEVD}] for momenta such that $(|M^{+}_{+1}|^2+|M^{+}_{-1}|^2)/2$ is at least $2\%$ of the maximum of (b). All spectra (b)--(d) are in atomic units and were obtained for a peak intensity $I \cpeq 10^{14}$~$\mathrm{W/cm^2}$ ($U_p=0.22$~a.u.), carrier frequency $\omega \cpeq 0.057$~a.u.\ ($\lambda \cpeq 800$~nm), $N=2$ cycles [\eqref{eq:vecA}] and ionization potential $I_p \cpeq 0.579$~a.u. of argon. All results were averaged over the carrier envelope phase $\delta$ and over orientations of the target corresponding to rotations of \eqref{eq:rhostate} around $z$. 
    }    
    \label{fig:diagrams}
\end{figure*}

Motivated by progress in the generation and measurement of free-electron vortices \cite{bliokh2017theory, asenjo2014dichroism, juchtmans_using_2015, maxwell2021manipulating, kang2021conservation, maxwell_entanglement_2022}---free-electrons with a helical phase front carrying orbital angular momentum (OAM)---here we propose an alternative and more direct approach to ultrafast imaging of molecular chirality, which does not rely on rotating electric fields. We propose to transfer chirality from the molecule and its initial electronic state directly
into the photoelectron wave packet with the help of an intense ultra-short \textit{linearly} polarized IR laser pulse.
The strong laser electric field lowers the barrier of the binding potential for a brief period of time, during
which the electron is released and accelerated away from the molecular ion. During this process, the chiral shape of the initial
electron wave function is imprinted in the three-dimensional phase structure of the released electron wave
packet and can be recovered via electron-OAM measurements [see FIG. \ref{fig:diagrams}(a)]. 

The helicity of the electron vortex is given by the projection of its linear momentum $\vec{p}$ on its OAM $\vec{l}$. Since this helicity is determined by the chirality of the initial wave function,  photoelectrons propagating in opposite directions have opposite OAM [see FIG \ref{fig:diagrams}(a)]. We call this effect PhotoElectron Vortex Dichroism (PEVD) and quantify it by direct integration of the time-dependent Schrödinger equation for a simple chiral target. Our numerical results show that PEVD is a very promising candidate for monitoring chiral dynamics with sub-femtosecond time resolution. 

\changesTwo{In the following we assume that the fixed nuclei approximation holds and that the vibrational degree of freedom plays a secondary role \cite{petretti_alignment-dependent_2010,sandor_strong_2016,muller_photoelectron_2020}. As usual in strong-field ionization, we will focus on the electronic wave function at the level of the single-active-electron approximation and considering molecular orientation as a parameter.}

To describe  strong-field ionization from our chiral target, we first employ the plane-wave-momentum transition amplitude
\begin{equation}
M(\mathbf{p}|\chi^{\epsilon})=\lim_{\substack{t_0\rightarrow-\infty\\t\rightarrow\infty}} \braket{\psi_\mathbf{p}(t)|U(t,t_0)|\chi^{\epsilon}(t_0)},
\label{eq:M_planewave}
\end{equation}
where 
$|\chi^{\epsilon}\rangle$ is the chiral initial state of the target, the superscript $\epsilon=\pm$ indicates the enantiomer,
$U(t,t_0)=\hat{T}\exp[\int_{t_0}^t d \tau \hat{H}(\tau)]$ is the time evolution operator (with $\hat{T}$ being the time-ordering operator) corresponding to the Hamiltonian $\hat{H}(\tau)=\hat{p}^2/2+V(\hat{r})+r_{||}E(\tau)$ in the single active electron and electric-dipole approximation,
and $\ket{\psi_\mathbf{p}}$ describes a continuum state with asymptotic momentum $\mathbf{p}$.
Atomic units are used throughout, unless stated otherwise. 

For the sake of both simplicity in the numerical implementation and clarity in the interpretation of the numerical results, we take the initial state to be the helical orbital shown in \figref{fig:diagrams}(a), which is given by a superposition of hydrogenic states according to \cite{ordonez2019propensity, ordonez_propensity_2022} \begin{align}
    \ket{\chi^{\epsilon}}&=\frac{\ket{\chi^{\epsilon}_1}+\ket{\chi^{\epsilon}_{-1}}}{\sqrt{2}} \quad \text{ with }
    \label{eq:rhostate}
    \\
    \ket{\chi^{\epsilon}_m}&=\frac{i\ket{4f_m}+ \sign(\epsilon m)\ket{4d_m}}{\sqrt{2}}.
    \label{eq:rho_part}
 \end{align}
Here, $\ket{n\ell_m}$ denotes the hydrogenic state with principal, orbital and magnetic quantum numbers $n$, $\ell$ and $m$, respectively.
\changesTwo{Orbitals with similar helical shapes are common among chiral allenes \cite{hhendon_helical_2013,bro-jorgensen_quantification_2021}, an important type of compound in organic chemistry  \cite{hhendon_helical_2013,neff_recent_2015}.
} 

We use an effective charge $Z=4\sqrt{2 I_p}$ to fix the ionization potential of $|\chi^\epsilon\rangle$ to that of argon $I_p=0.579$ a.u., a well studied target in strong field ionization. The enantiomers $|\chi^+\rangle$ and $|\chi^-\rangle$ are related to each other by reflection in the $xz$ plane.
The decomposition of $|\chi^\epsilon\rangle$ into its azimuthal angular momentum components $|\chi^\epsilon_{m=\pm1}\rangle$ in Eq. (\ref{eq:rho_part}) reveals an enantio-sensitive correlation \footnote{For $\epsilon=+$, $m$ and $\mathrm{sign}(\epsilon m)$ have the same sign. For $\epsilon=-$, $m$ and $\mathrm{sign}(\epsilon m)$ have opposite signs.} between the sign of $m$ (rotation around $z$) and the relative phase between the $d$ and $f$ components (motion/asymmetry along $z$) in $|\chi_{\pm1}^{\epsilon}\rangle$. 
Such a correlation between rotation and motion along the rotation axis is not a peculiarity of these states, but rather a quintessential feature of chiral objects (e.g. a corkscrew). As we discuss below, this general feature is at the heart of PEVD.

In order to model the photoelectron OAM in strong field ionization we will expand the transition amplitude in terms of vortex states with a well-defined OAM $l_v$, these take the form \cite{bliokh2017theory}
\begin{align}
\psi_{l_v}(\mathbf{r})\propto J_{l_v} (p_{\bot}r_{\bot}) e^{i{l_v}\phi_r} e^{ip_\| r_\|},
\label{eq:vortex_state}
\end{align}
where $J_{l_v}(p_{\bot}r_{\bot})$ is the Bessel function of the first kind
and ($r_\perp, r_\parallel, \phi_r$) and ($p_\perp, p_\parallel, \phi_p$) are the cylindrical coordinates of $\mathbf{r}$ and $\mathbf{p}$, respectively.
The OAM-resolved transition amplitude, in terms of the transition amplitude [\eqref{eq:M_planewave}], yields
\begin{align}
    M_{l_v}(p_\|,p_\bot|\chi^{\epsilon})&=\frac{i^{l_v}}{2\pi}\int_{-\pi}^{\pi} d\phi_p \hspace{2pt} e^{-i{l_v}\phi_p} M(\mathbf{p}|\chi^{\epsilon})
     \label{eq:M_vortex}
    \\
    &=i^{l_v} M(\mathbf{p}|\chi^{\epsilon}_{l_v})|_{\phi_p=0}. \notag
\end{align}
The final line exploits the conservation of azimuthal angular momentum, when using a linear field aligned as in \figref{fig:diagrams}(a), so that the photoelectron OAM will only include contributions from the bound state with $m=l_v$. In the following, we use the shorthand notation $M^{\epsilon}\equiv M(p_\parallel,p_\perp|\chi^\epsilon)$ for the plane-wave amplitude and  $M_{l_{v}}^{\epsilon}\equiv M_{l_{v}}(p_\parallel,p_\perp|\chi^\epsilon)$ for the OAM-resolved amplitude.

We will employ a 2-cycle laser field linearly polarized along the $z$-axis, see \figref{fig:diagrams}(a). PEVD also occurs for longer pulses, but the interpretation is simplest for short pulses, where re-collision of the electron with the parent ion is suppressed. The electric field is given by $\mathbf{E}(t)=-\partial_t \mathbf{A}(t)$ and the vector potential can be written as
\begin{equation}
\mathbf{A}(t)=A_0 \sin^2\left(\frac{\omega t}{2N}\right)\cos\left(\omega t+\delta\right) \mathbf{\hat{e}}_z.
\label{eq:vecA}
\end{equation}
Here, $A_0=2\sqrt{U_p}$, $U_p$ is the ponderomotive energy, $\omega$ the laser frequency, $N$ the number of laser cycles and $\delta$ the carrier-envelope phase (CEP). In order to avoid asymmetries in the angular distributions resulting from the laser field, all results will be averaged over the CEP $\delta$.

We use \qprop\ to compute the OAM-resolved momentum distribution via \eqref{eq:M_vortex}. \qprop\ \cite{tulsky2020qprop,*mosert2016photoelectron} enables the efficient simulation of a single active electron bound in a spherically symmetric potential, interacting with an intense laser field. As previously stated, we utilize a Coulomb potential with an effective charge $Z=4\sqrt{2 I_p}$, so that the binding energy is equal to $-I_p$. 

In \figref{fig:diagrams}, we demonstrate the enantio-sensitive asymmetry present in PEVD. Panel (a) summarizes the key finding of this work, namely that photoionization of a chiral target with linearly polarized light yields photoelectron vortices whose helicity encodes the handedness of the chiral target. In terms of OAM, this means that photoelectrons emitted in opposite directions carry opposite OAM. 
This effect is quantified in the remaining panels, which display CEP-averaged momentum distributions, computed using \qprop\ for an aligned \footnote{We average over all target orientations resulting from a rotation of \eqref{eq:rhostate} around the $z$ axis followed by a rotation by either $0$ or $\pi$ radians around the $y$ axis.} target $\ket{\chi^{+}}$.
For the sake of comparison, panel (b) shows the usual plane-wave photoelectron momentum distribution [see \eqref{eq:M_planewave}]. 
The signal, which is mostly located near the polarization axis, is completely symmetric 
with respect to $p_{\parallel}$ and is not enantio-sensitive.
Panel (c) shows the OAM-resolved momentum distribution [\eqref{eq:M_vortex}] for OAM $l_v=+1$ ($|M_{+1}^{+}|^2$) and $l_v=-1$ ($|M_{-1}^{+}|^2$), respectively.
Now the emission displays an OAM-dependent asymmetry with respect to $p_\parallel$, 
such that photoelectrons with positive [negative] OAM are emitted preferentially in the negative [positive] $p_\parallel$ direction.

To quantify this OAM dependence we proceed as usual in other chiro-optical techniques and define PEVD as the difference between the signals corresponding to opposite OAM normalized by their average,
\begin{equation}
    \mathrm{PEVD}_{l_v}^{\epsilon} \equiv 2\frac{\left|M_{l_{v}}^{\epsilon}\right|^{2}-\left|M_{-l_{v}}^{\epsilon}\right|^{2}}{\left|M_{l_{v}}^{\epsilon}\right|^{2}+\left|M_{-l_{v}}^{\epsilon}\right|^{2}}=2\frac{\left|M_{l_{v}}^{\epsilon}\right|^{2}-\left|M_{l_{v}}^{-\epsilon}\right|^{2}}{\left|M_{l_{v}}^{\epsilon}\right|^{2}+\left|M_{l_{v}}^{-\epsilon}\right|^{2}}.
    \label{eq:PEVD}
\end{equation}
The second equality, which explicitly shows the enantio-sensitive character of this measure ($\mathrm{PEVD}_{l_v}^{-\epsilon}=-\mathrm{PEVD}_{l_v}^{\epsilon}$), follows from the property $|M_{-l_{v}}^{-\epsilon}|^2=|M_{l_{v}}^{\epsilon}|^2$, which we have verified in our calculations and is required by symmetry. Indeed, mirror reflection of the setup shown in \figref{fig:diagrams}(a), in the $xz$ plane, inverts the enantiomer as well as the OAM of each vortex.
In panel (d) we can see that $\mathrm{PEVD}_{1}^{+}$ reaches values approaching the theoretical extremes of $\pm200\%$, showing that it is an extremely sensitive measure of molecular chirality.  \changes{Note that (i) we made no attempt to maximize this value by tweaking the initial wave function and (ii) an actual molecular electronic state will contain, in general, several $m$ components, leading to several $l_v$ values of the photoelectron OAM, and thus to several $\mathrm{PEVD}_{l_v}^{\epsilon}$ distributions, i.e. multiple enantio-sensitive observables.}

The $p_\parallel$ asymmetry in panel (c) derives from conservation of OAM, which enforces that only the bound-state component $\ket{\chi^{+}_{m}}$ [\eqref{eq:rho_part}] with $m=l_v$ contributes to photoelectrons with OAM $l_v$ [\eqref{eq:M_vortex}]. 
The $z$ asymmetry of this single-OAM component $|\chi_{l_v}^{+}\rangle$ is imprinted on the photoelectron vortex momentum distribution $|M_{l_v}^{+}|^2$.
Because of the chiral correlation between rotation around $z$ and motion/asymmetry along $z$ in $|\chi_{m}^{\epsilon}\rangle$, 
the $z$ asymmetry of $|\chi_{l_v}^{+}\rangle$ is opposite for opposite values of $l_v$ [\eqref{eq:rho_part}], 
and thus
$|M_{l_v}^{+}|^2$ and $|M_{-l_v}^{+}|^2$ display opposite asymmetries with respect to $p_{\parallel}$.

\begin{figure*}
    \centering
    \includegraphics[width=\textwidth]{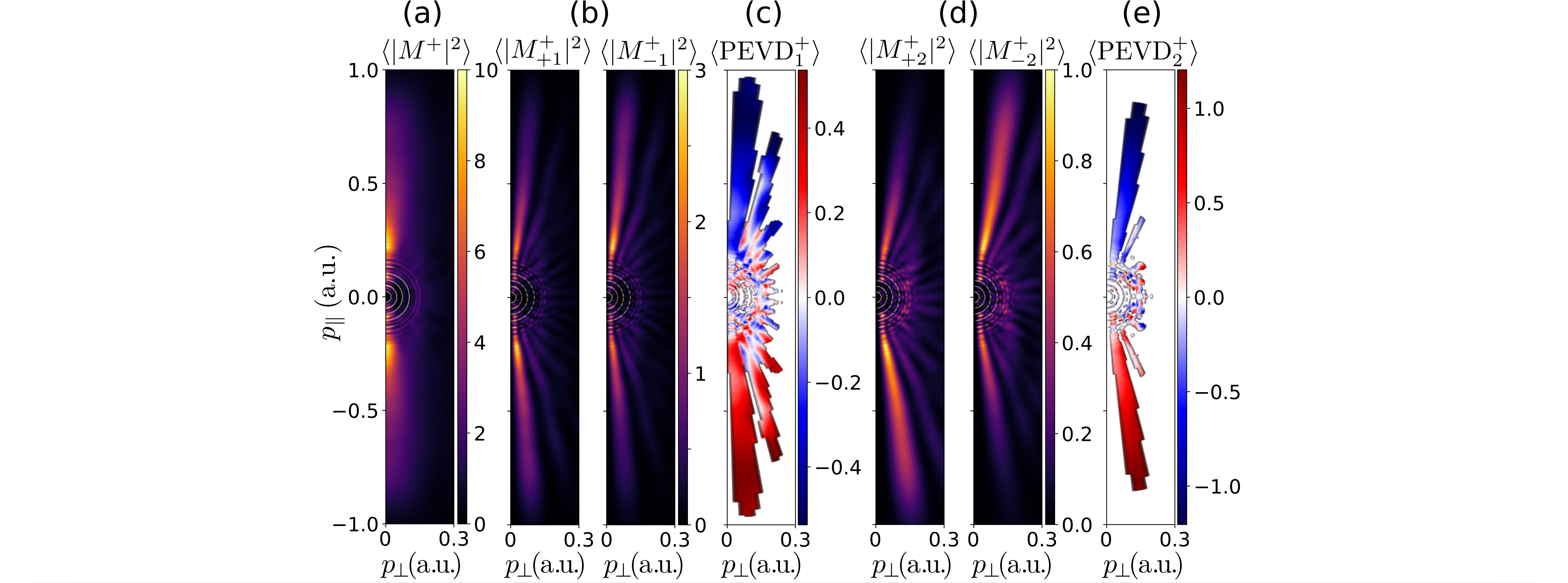}
    \caption{Photoelectron spectra as in \figref{fig:diagrams} but for a partially aligned chiral target with a distribution of orientations given by $P(\beta)=3\cos^2(\beta)$  ($\langle \cos^2\beta\rangle=0.6$), see the supplemental material \cite{supplemental_material} for details.
    }
    \label{fig:Qprop}
\end{figure*}

While virtually perfect molecular alignment is possible \cite{holmegaard_laser-induced_2009} and we expect PEVD to be maximal in this case, we consider also PEVD for a modest degree of alignment.
The inclusion of different orientations leads to contributions of orbitals with $m\in [-3,3]$, and correspondingly, $l_v \in [-3,3]$. For more details on orientation averaging, see the supplemental material \cite{supplemental_material}.
The orientation-averaged CEP-averaged momentum distributions
calculated with \qprop\ are shown in \figref{fig:Qprop}. The distribution $P(\beta)=3\cos^2(\beta)$ leads to target orientation such that, $\langle\cos^2(\beta)\rangle=0.6$,
which is routinely achievable in experiments by an alignment pre-pulse \cite{stapelfeldt2003colloquium}. 
Note that $\beta$ is the angle between the $z$ axes of the molecular and laboratory frames \cite{brink_angular_1968}.
As before, in panel (a) we plot the plane-wave momentum distribution, which remains symmetric, but is now averaged over the molecular orientations, introducing orbitals with additional $m$ values, in particular $m=0$, that leads to a strong signal along the $p_{\parallel}$ axis.
In panels (b) and (d), the OAM-resolved momentum distribution is plotted for $l_v=\pm1$ and $l_v=\pm2$, respectively. Panel (b) is equivalent to \figref{fig:diagrams}(c), the partial alignment leads to some reduction in the asymmetry, such that $|\langle$PEVD$^{+}_{1}\rangle|$, panel (c), peaks at $50\%$. On the other hand, the higher OAM $l_v=\pm2$ in panel (d) allows for a larger asymmetry, as the initial state $\ket{\chi^{\pm}_2}$ exhibits even more bias in the momentum distribution, enabling $|\langle$PEVD$^{+}_{2}\rangle|\approx 120\%$ to be achieved in panel (e). 
The signal of $\langle|M^{+}_{\pm2}|^2\rangle$ is lower than that of $\langle|M^{+}_{\pm1}|^2\rangle$, but a sizable region in momentum is larger than $2\%$ of the peak of $\langle|M^{+}|^2\rangle$, and thus could be reasonably measured.
We do not plot $\langle|M^{+}_0|^2\rangle$ or $\langle|M^{+}_{\pm3}|^2\rangle$ as they are both symmetric and not enantio-sensitive, which is because the photoelectron will derive from a single symmetrical $s$- and $f$-state, respectively.

Like most chiral effects \cite{berova_comprehensive_2012,ordonez_chiral_2019,janssen_detecting_2013,fischer_nonlinear_2005,patterson_sensitive_2013,ayuso_synthetic_2019,asenjo2014dichroism}, we expect PEVD to occur even for randomly oriented samples. However, reproducing such behavior with photoelectrons requires including the effect of the anisotropy of the 
molecular potential on the continuum states\changes{, which can be interpreted as a higher order contribution to the continuum states (see e.g. \cite{dreissigacker2014photoelectron}).}

We have also confirmed 
that the enantio-sensitive asymmetry may be replicated by the strong field approximation (SFA), see supplementary material for details \cite{supplemental_material}. The SFA uses an intuitive quantum orbit formalism, see e.g. Refs.~\cite{keldysh1965ionization,*faisal1973multiple,*reiss1980effect,becker2002above,amini2019symphony}. Within the SFA the asymmetry can also be partially attributed to interference between wave packets that  tunnel ionize from either side of the target. For the short pulses used here such interference does not play a role, but it is relevant for longer laser pulses, where it may allow for holographic chiral imaging. \changes{In the supplemental material Fig.~1 we show PEVD for a 10-cycle pulse, where rich features due to recollision can be observed. The detailed mechanisms behind these features} could be investigated by a `Coulomb-corrected' quantum-orbit model such as the Coulomb quantum-orbit strong field approximation \cite{maxwell2017coulombcorrected,maxwell2018coulombfree,maxwell2018analytic}.

In summary, we have showcased the newly discovered effect---photoelectron vortex dichroism (PEVD), which leads to a large asymmetry in the orbital angular momentum (OAM)-resolved photoelectron momentum distributions for a chiral target subjected to a strong linearly polarized laser field. This asymmetry is the manifestation of a laser-assisted transfer of chirality from the bound orbital to the photoelectron vortex wavepacket, which provides an intrinsically chiral observable. \changes{The emergence of PEVD reflects the fact that chiral molecules are a natural source of electron vortices that relies neither on the spin angular momentum (SAM) of the photon nor on the net OAM of the initial state}.
The high enantio-sensitivity of PEVD was demonstrated numerically using the time-dependent Schr\"odinger equation solver \qprop\ \cite{tulsky2020qprop,*mosert2016photoelectron}, for aligned and partially aligned targets. 
The enantio-sensitivity of the OAM asymmetry in PEVD is most clearly visible after averaging out the CEP asymmetry induced by the laser pulse and  
provides a 
robust experimental observable.
 Similarly to other chiral effects such as photoelectron circular dichroism (PECD) \cite{beaulieu_universality_2016}, we expect PEVD to occur also in randomly oriented samples and furthermore to be a `universal' effect occurring across all photoionization regimes.

In contrast to most electric-dipole chiral effects \footnote{The only (partial) exception we are aware of is Ref. \cite{ayuso_ultrafast_2021}, where linear polarization is used but nevertheless turned into elliptical polarization via tight focusing in the interaction region.}, which rely on circular \cite{ritchie1976theory} or on combinations of several linearly polarized fields \cite{fischer_nonlinear_2005, patterson_sensitive_2013, demekhin_photoelectron_2018, ayuso_synthetic_2019}, PEVD relies only on a single beam of linearly polarized light, which is possible thanks to the chiral character of the observable---the OAM helicity \footnote{The OAM is a pseudovector, which with the propagation direction, provides an inherently chiral observable, in contrast to the vector character of the usual electric-dipole chiral observables \cite{ordonez2018generalized}.}. 
This is an important experimental advantage, in particular when using (broadband and short-wavelength) atto-second pulses, where we also expect PEVD to occur and where the polarization control required by other chiral effects is complicated by lossy and dispersive optics \cite{huang_polarization_2018}. \changes{More importantly, 
in the strong-field regime linearly polarized light allows for recollision, 
which leads to processes
such as high-harmonic generation \cite{lewenstein_theory_1994}, high-order above-threshold ionization \cite{becker2002above}, non-sequential double ionization \cite{demorissonfaria_electron_2011,becker_theories_2012}, 
laser-induced electron diffraction \cite{zuo_laserinduced_1996} and photoelectron holography \cite{figueirademorissonfaria_it_2020}. 
Thus, PEVD 
sets the stage for the study of
the role of the electron OAM and the opportunities it offers in 
recollision-based 
ultrafast chiral imaging.} 

\changes{Another key aspect of PEVD is that while the SAM of the photon is strictly limited to $-1$ and $1$, the OAM of the photoelectron is limited only by the complexity of the initial wave function and can take any integer value, 
thus yielding a rich array of enantio-sensitive observables.}

To observe PEVD in experiment requires, in the simplest case, the counting of photoelectrons travelling in one direction along the laser field polarization axis and measurement of the OAM (and thus the associated helicity).
\changes{A lot of activity in the
field of electron optics has resulted in an increasing variety of methods (inspired by their
analogues in optics) for the measurement of the OAM of an electron wave. These range from simply diffracting the electron wave through a knife edge \cite{guzzinati_measuring_2014}, to mode conversion with astigmatic lenses \cite{schattschneider_novel_2012}, and very effective state-of-the-art OAM sorting using conformal mapping \cite{grillo_measuring_2017, tavabi_experimental_2021}, among others (see Refs. \cite{lloyd2017electron,bliokh2017theory} for reviews).
This variety of methods have built upon the electron optics found in transmission electron
microscopes (TEMs), and thus we propose to observe PEVD by using chiral molecules as a laser-triggered electron source \cite{baum_physics_2013}
in a TEM already fitted with an OAM measurement stage (see supplementary material for details)}. 

\begin{acknowledgments}
The data and plotting scripts for all figures used in this manuscript is freely available on a \href{https://doi.org/10.5281/zenodo.6939947}{Zenodo database}. The code written for this manuscript has been made open source and is available as \href{https://zenodo.org/badge/latestdoi/454731109}{a release on Zenodo} and the repository is on \href{https://github.com/asmaxwell/SFA_OAM_Linear}{GitHub}.

The authors would like to thank Dr. Allan Johnson for enlightening discussions about polarization control and Prof. Lars Bojer Madsen for illuminating discussions, careful reading of the draft and for pointing out some key references.
ASM acknowledges funding support from the European Union’s Horizon 2020 research and innovation programme under the Marie Sk\l odowska-Curie grant agreement, SSFI No.\ 887153.
XBP, AO, ML and ASM acknowledge support from 
ERC AdG NOQIA; Ministerio de Ciencia e Inovación, Agencia Estatal de Investigaciones (PGC2018-097027-B-I00/10.13039/501100011033,  CEX2019-000910-S/10.13039/501100011033, Plan Nacional FIDEUA PID2019-106901GB-I00, FPI, QUANTERA MAQS PCI2019-111828-2, QUANTERA DYNAMITE PCI2022-132919,  Proyectos de I+D+I “Retos Colaboración” QUSPIN RTC2019-007196-7); European Union NextGenerationEU (PRTR);  Fundació Cellex; Fundació Mir-Puig; Generalitat de Catalunya (European Social Fund FEDER and CERCA program AGAUR Grant No. 2017 SGR 134, QuantumCAT \ U16-011424, co-funded by ERDF Operational Program of Catalonia 2014-2020); Barcelona Supercomputing Center MareNostrum (FI-2022-1-0042); EU Horizon 2020 FET-OPEN OPTOlogic (Grant No 899794); National Science Centre, Poland (Symfonia Grant No. 2016/20/W/ST4/00314); European Union’s Horizon 2020 research and innovation programme under the Marie-Skłodowska-Curie grant agreement No 101029393 (STREDCH) and No 847648  (“La Caixa” Junior Leaders fellowships ID100010434: LCF/BQ/PI19/11690013, LCF/BQ/PI20/11760031,  LCF/BQ/PR20/11770012, LCF/BQ/PR21/11840013).
\end{acknowledgments}

\bibliographystyle{apsrev4-1}
\bibliography{bibliographyPRL}

\newpage


\section*{Supplementary material}

# Ultrafast imaging of molecular chirality with photoelectron vortices - SUPPLEMENTAL MATERIAL -

Xavier Barcons Planas,[1] Andrés Ordóñez,[1] Maciej Lewenstein,[1, 2] and Andrew S. Maxwell[1, 3]

[1]*ICFO-Institut de Ciencies Fotoniques, The Barcelona Institute of Science and Technology, 08860 Castelldefels (Barcelona), Spain*
[2]*ICREA, Pg. Lluís Companys 23, 08010 Barcelona, Spain*
[3]*Department of Physics and Astronomy, Aarhus University, DK-8000 Aarhus C, Denmark*

PEVD is briefly examined for a longer 10-cycle pulse. Furthermore, we reproduce the PEVD effect using the strong field approximation (SFA), provide information on the SFA formalism, and an analysis of the effect of rotating the target. The derivation of the length gauge prefactor for hydrogenic states, in the saddle point approximation without divergences, is provided. The target is rotated to account for a partially aligned sample, and the influence of the target rotation on the transition amplitude is supplied. We also provide a discussion about the experimental implementation, limitations and potential solutions.

## PEVD FOR A 10-CYCLE LASER PULSE

In the main manuscript, we used a 2-cycle pulse to demonstrate photoelectron vortex dichroism (PEVD), see FIG. 1 therein. The motivation for this was that the analysis is simpler. However, this does not mean that for longer laser envelopes the effect is reduced. In fact, it can be argued that there are richer dynamics and more information to be gleaned when using longer pulses, due to the prominent role that will be played by recolliding electrons. In FIG. 1 of this supplemental, we show the same distributions as in FIG. 1 of the main manuscript, but now for a 10-cycle pulse. The normalised difference, $PEVD_1^+$ (FIG. 1(c)), is just as high as before, but now there are regions, for which the difference flips sign. The highest energy regions are actually the same sign as before, however, at around $2U_p$ in energy a sign flip occurs. This may be due to 'softly recolliding' low energy trajectories [1–3], which undergo more than one laser driven return before scattering off the target. This would mean that an electron wavepacket that tunnels on a particular side of the chiral target (thus having a particular OAM) ends up going in the opposite direction. In a lower energy region, at just under $|p_\parallel| \approx 0.2$, the sign flips again, suggesting further laser-driven returns, which is consistent with the behaviour of the low-energy structure [4, 5].

In summary, for longer laser pulses PEVD survives, in fact it is richer, with recollisions playing a greater role. A complete description is beyond the scope of this work and would require a trajectory analysis, which could provide a connection to imaging tools such as laser-induced electron diffraction (LIED) [6] and photoelectron holography [7].

## THE SFA AND BOUND-STATE PREFACTOR

The evolution of an electron under the influence of a binding potential $V$ and a strong laser field can be described by a scattering matrix transition amplitude

$$M_f = \lim_{\substack{t \to \infty \\ t_0 \to -\infty}} \langle \psi_f(t) | U(t, t_0) | \psi_0(t_0) \rangle, \tag{1}$$

which in the SFA, for a final continuum state $|\psi_{\mathbf{p}}\rangle$ with asymptotic momentum $\mathbf{p}$, reduces to [8–11]

$$M(\mathbf{p}) = -i \int_{-\infty}^{+\infty} dt' d(\mathbf{p}, t') e^{iS(\mathbf{p}, t')}, \tag{2}$$

where the integral can be computed using the saddle point approximation as a sum over the different saddle times $t_s$,

$$M(\mathbf{p}) \approx \sum_{t_s} \sqrt{\frac{2\pi i}{S''(\mathbf{p}, t_s)}} d(\mathbf{p}, t_s) e^{iS(\mathbf{p}, t_s)}. \tag{3}$$

The exponential terms from the continuum evolution and bound state evolution are collected together into a quasi-classical action, $S(\mathbf{p}, t) = I_p t + \int_{-\infty}^{t} d\tau (\mathbf{p} + \mathbf{A}(\tau))^2 / 2$, which describes the propagation of an electron from the ionization time to the end of the pulse. $I_p$ is the ionization potential, $\mathbf{p}$ the electron's momentum, $\mathbf{A}(\tau)$ the vector

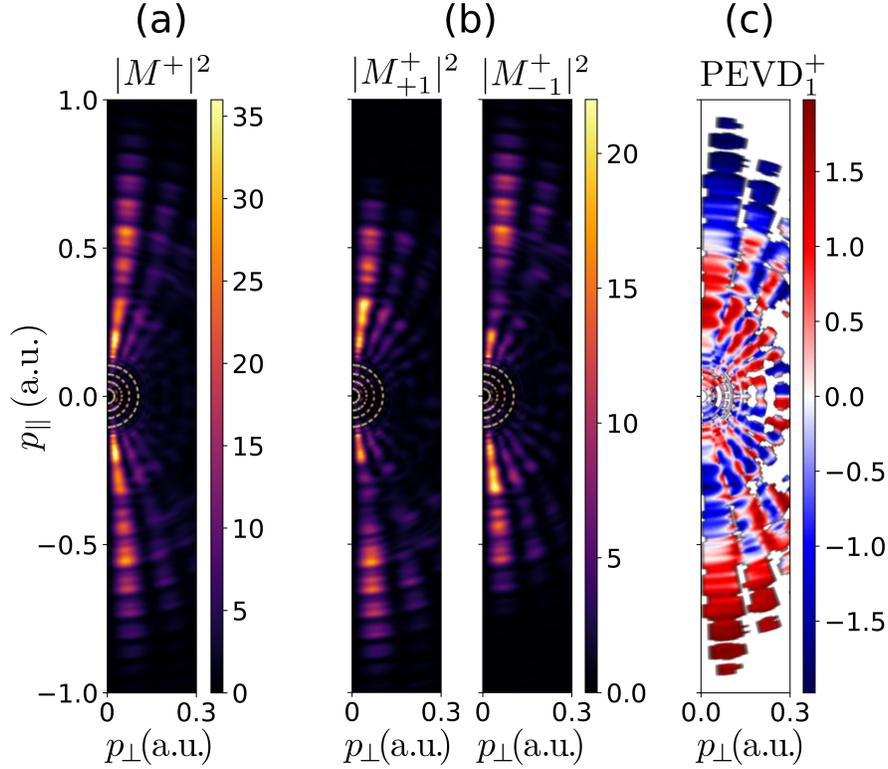

FIG. 1: Equivalent to the results shown in Fig. 1 of the main manuscript, except a 10-cycle pulse envelope is used. (a) Standard photoelectron spectrum resulting from strong-field ionization of a perfectly aligned chiral target $|\chi^+\rangle$ as a function of the momentum perpendicular and parallel to the laser field. (b) OAM-resolved photoelectron spectra for OAM $l_v = +1$ (left) and $l_v = -1$ (right). (c) Difference of the OAM-resolved spectra normalized by their average for momenta such that $(|M^+_{+1}|^2 + |M^+_{-1}|^2)/2$ is at least 2% of the maximum of (b). All spectra (b)–(d) are in atomic units and were obtained for a peak intensity $I = 10^{14}$ W/cm² ($U_p = 0.22$ a.u.), carrier frequency $\omega = 0.057$ a.u. ($\lambda = 800$ nm), $N = 10$ cycles and ionization potential $I_p = 0.579$ a.u. of argon. All results were averaged over the carrier envelope phase $\delta$ and over orientations of the target corresponding to rotations of $|\chi^+\rangle$ around $z$.

potential and the ionization times $t_s$ are solutions to stationary action $\partial_t S(\mathbf{p}, t_s) = I_p + \frac{1}{2}(\mathbf{p} + \mathbf{A}(t_s))^2 = 0$, see [12] for more details on saddle-point solutions. Furthermore, the information about the interaction and electronic bound state is considered in the bound-state prefactor $d(\mathbf{p}, t') = \langle \tilde{\mathbf{p}}(t') | V(\hat{\mathbf{r}}) | \psi_0 \rangle$. For a hydrogenic state $\psi_0$ with $n$, $\ell$ and $m$ quantum numbers the prefactor is given by

$$d(\mathbf{p}, t') = \alpha_{n\ell} Y_\ell^m (\theta_{\tilde{p}}, \phi_{\tilde{p}}) \sum_{k=0}^{n-\ell-1} \beta_{n\ell k} \tilde{\mathbf{p}}^\ell(t') {}_2F_1(a, b; c; z), \qquad (4)$$

where $Y_\ell^m$ is the spherical harmonic function, ${}_2F_1(a, b; c; z)$ is the ordinary hypergeometric function and $\tilde{\mathbf{p}}(t) = \mathbf{p}$ [$\tilde{\mathbf{p}}(t) = \mathbf{p} + \mathbf{A}(t)$] in the velocity [length] gauge. Here, $\alpha_{n\ell} = (-i)^\ell \sqrt{\frac{(n-\ell-1)!}{2n(n+\ell)!}}$, $\beta_{n\ell k} = (-2)^k \frac{(n+\ell)!(\sqrt{2I_p})^{-\frac{1}{2}-\ell}}{\Gamma(3/2+\ell)(n-\ell-k-1)!k!}$, $a = (2 + k + 2\ell)/2$, $b = (3 + k + 2\ell)/2$, $c = 3/2 + \ell$ and $z = -\tilde{\mathbf{p}}^2(t')/(2I_p)$.

In the length gauge, $d(\mathbf{p}, t_s) \to \infty$, since $\tilde{\mathbf{p}}^2(t_s) = -2I_p$. The divergence is solved as follows. First, we write the ${}_2F_1$ hypergeometric function as a power series and do the variable transformation $F(a, b; c; z) = (1-z)^{-a} F(a, c-b; c; z/(z-1))$ [13]. Then, Eq. (2) becomes

$$M(\mathbf{p}) = \sum_{k=0}^{n-\ell-1} \sum_{\nu=0}^{\infty} \underbrace{\int_{-\infty}^{+\infty} dt' \frac{1}{[\tilde{\mathbf{p}}^2 + 2I_p]^{\nu+a}} e^{iS(\mathbf{p}, t')} \nu_{n\ell m k \nu}(\mathbf{p}, t')}_{I_{sp}}, \qquad (5)$$

where $\nu_{n\ell m k\nu}(\mathbf{p}, t') = \frac{\Gamma(c)}{\Gamma(a)\Gamma(c-b)} \frac{\Gamma(a+\nu)\Gamma(c-b+\nu)}{\Gamma(c+\nu)\nu!} (2I_p)^a \alpha_{n\ell} Y_\ell^m (\theta_{\tilde{p}}, \phi_{\tilde{p}}) \beta_{n\ell k} \tilde{\mathbf{p}}^{\ell+2\nu}$. However, when applying the saddle-point method to solve the integral $I_{sp}$, we still have a singularity. This can be solved by exploiting the quadratic

expansion of the action around saddle-point [14]. In the vicinity of the saddle point $t_s$, $S'(t_s) = 0$, we have $S'(t) \approx S''(t_s)(t - t_s)$. Therefore, since in the saddle-point approximation $\tilde{\mathbf{p}}^2 + 2I_p = S'(t)$,

$$I_{sp} = \frac{\nu_{n\ell mk\nu}(\mathbf{p}, t_s)}{[S''(\mathbf{p}, t_s)]^{\nu+a}} \underbrace{\int_{-\infty}^{+\infty} dt' \frac{1}{(t' - t_s)^{\nu+a}} e^{iS(\mathbf{p}, t')}}_{I_2}.  \tag{6}$$

To solve the $t' = t_s$ singularity of $I_2$, we use the transformation $\frac{1}{(x - x_0)^\eta} = \frac{1}{\Gamma(\eta)} \int_0^\infty d\xi \, \xi^{\eta-1} \exp\left[-\xi(x - x_0)\right]$, and solving both integrals we get

$$I_2 = i^{\frac{\nu+a}{2}} \frac{\Gamma(\frac{\nu+a}{2})}{2\Gamma(\nu+a)} \sqrt{\frac{2\pi i}{S''(\mathbf{p}, t_s)}} [2S''(\mathbf{p}, t_s)]^{\frac{\nu+a}{2}} e^{iS(\mathbf{p}, t_s)}.  \tag{7}$$

Since $I_{sp} = \nu_{n\ell mk\nu}(\mathbf{p}, t_s)[S''(\mathbf{p}, t_s)]^{-(\nu+a)} I_2$, Eq. (5) becomes

$$M(\mathbf{p}) = \frac{1}{2} \sqrt{\frac{2\pi i}{S''(\mathbf{p}, t_s)}} e^{iS(\mathbf{p}, t_s)} \tilde{d}(\mathbf{p}, t_s),  \tag{8}$$

with prefactor

$$\tilde{d}(\mathbf{p}, t_s) = \sum_{k=0}^{n-\ell-1} \sum_{\nu=0}^{\infty} \nu_{n\ell mk\nu}(\mathbf{p}, t_s) i^{\frac{\nu+a}{2}} \frac{\Gamma(\frac{\nu+a}{2})}{\Gamma(\nu+a)} 2^{\frac{\nu+a}{2}} [S''(\mathbf{p}, t_s)]^{-\frac{\nu+a}{2}}.  \tag{9}$$

Solving the previous sum over $\nu$ writing it as a combination of $_3F_3$ hypergeometric functions and replacing all the values, we get the expression of the bound-state prefactor in the length gauge for any hydrogenic state without divergences:

$$\begin{aligned}
\tilde{d}(\mathbf{p}, t_s) &= (-i)^\ell \sqrt{\frac{(n-\ell-1)!}{2n(n+\ell)!}} Y_\ell^m(\theta_{\tilde{p}}, \phi_{\tilde{p}}) \sum_{k=0}^{n-\ell-1} \frac{(-1)^k (n+\ell)! 2^k (2I_p)^{\frac{3}{4} + \frac{\ell}{2} + \frac{k}{2}} \tilde{\mathbf{p}}^\ell}{(n-\ell-k-1)! \, k! \, (i\tilde{\mathbf{p}} \cdot \mathbf{E}(t_s))^{\frac{2+k+2\ell}{4}}} \\
&\quad \times \frac{1}{\Gamma(\frac{3}{2} + \ell) \Gamma(\frac{2+k+2\ell}{2})} \left[ \Gamma(\frac{2+k+2\ell}{4}) F_1 - k \frac{\sqrt{4I_p^2}}{i\tilde{\mathbf{p}} \cdot \mathbf{E}(t_s)} \frac{\Gamma(\frac{4+k+2\ell}{4})}{3+2\ell} F_2 \right],
\end{aligned}  \tag{10}$$

where $F_1 = {}_3F_3\left(\frac{2+k+2\ell}{4}, \frac{-k}{4}, \frac{2-k}{4}; \frac{1}{2}, \frac{5+2\ell}{4}, \frac{3+2\ell}{4}; \frac{I_p^2}{i\tilde{\mathbf{p}} \cdot \mathbf{E}(t_s)}\right)$ and $F_2 = {}_3F_3\left(\frac{4+k+2\ell}{4}, \frac{2-k}{4}, \frac{4-k}{4}; \frac{3}{2}, \frac{5+2\ell}{4}, \frac{7+2\ell}{4}; \frac{I_p^2}{i\tilde{\mathbf{p}} \cdot \mathbf{E}(t_s)}\right)$.

## PEVD IN THE SFA

The direct SFA gives a rough approximation to strong field ionization with a linearly polarized field, where fundamental coarse-grain features can be qualitatively reproduced in terms of a simple and intuitive picture. The transition amplitude may be computed using Eq. (3). FIG. 2 shows that the SFA can reproduce PEVD in both the velocity [top row] and length gauges [bottom row]. We verified the reversal of the asymmetries for the opposite enantiomer. Furthermore, comparing FIGs. 2(c) with FIG. 1(d) in the main text reveals that both gauges roughly coincide with the TDSE results in sign and magnitude. The difference in the shapes of the distributions are due to the lack of Coulomb focusing or Coulomb phases [7] in the SFA.

In the velocity gauge, the term $\langle \tilde{\mathbf{p}}(t_s) | V | \chi_{l_v}^\epsilon \rangle = \langle \mathbf{p} | V | \chi_{l_v}^\epsilon \rangle$ carrying the enantiomeric dependence can be factored out of the sum in Eq. (3). Thus, (i) the $p_\parallel$-asymmetry of $M(\mathbf{p} | \chi_{l_v}^\epsilon)$ in FIG. 2(b) can be directly traced back to the $z$-asymmetry of $|\chi_{l_v}^\epsilon\rangle$. And (ii) the interference fringes—visible in FIG. 2(a) and (b) [top row], which are due to photoelectron wave packets released in different half cycles and initially travelling in opposite directions—do not encode the target's chirality. Hence, the interference is not visible in the PEVD$_1^+$ normalized difference, panel (c) [top row]. This means that, the asymmetry, and in particular PEVD$_1^+$, is dependent only on the final momentum and is independent of the laser field. In the TDSE there is some dependence of the asymmetry on the laser field, so an additional mechanism must be involved.

In the length gauge SFA tunneling picture, the interpretation is more nuanced. Like the velocity gauge, there is a fixed $p_\parallel$ asymmetry, which changes sign with the OAM and enantiomer, however, this effect is smaller and does not completely account for the asymmetry observed in the TDSE with QPROP. In this case, the bound-state matrix

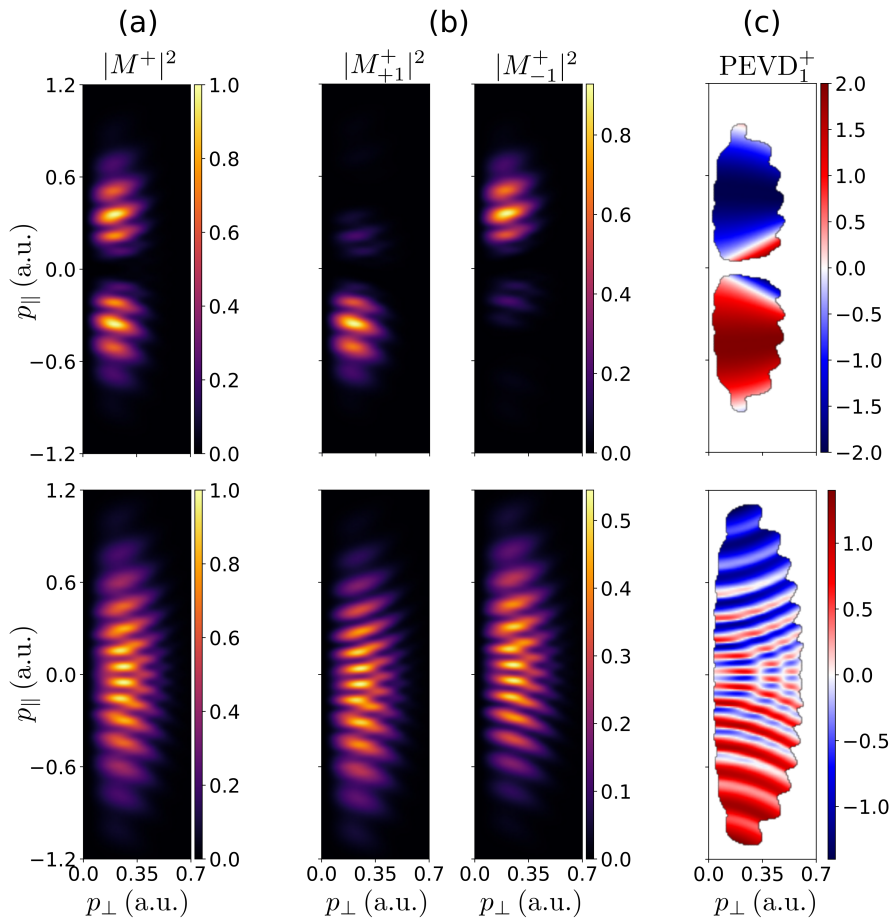

FIG. 2: Photoelectron spectra, as in Fig. 1 in the main text, but calculated using the SFA, with Eq. (3), in the velocity gauge (top row) and in the length gauge (bottom row). The spectra have been normalized to the maximum of $|M^+|^2$ for each gauge.

element depends on the time of ionization and the direction that the photoelectron wave packet tunnels out. We have verified that this leads to phases differences accumulated by photoelectron wave packets leaving from opposite sides of the target, which change sign with the OAM and enantiomer. The interference of these wave packets leads to enantio-sensitive interference fringes, which can clearly be seen in the $\mathrm{PEVD}_1^+$ normalized difference plot FIG. 2(c) [lower row]. These fringes actually increase the normalized difference peak from 100% to 150%. The presence of interferences encoding chiral information may play a larger role for longer pulses, such as those shown in FIG. 1, and could be used in an OAM-enhanced form of photoelectron holography [7, 15] to distinguish chiral enantiomers.

The length and velocity gauge SFA are known to disagree to some extent [16] due to the approximations made upon the continuum states. So the agreement of an OAM and enantiomer sensitive asymmetry demonstrates that this is a fundamental feature, predicted by all models. However, the interference predicted by the length gauge allows field parameters to affect the asymmetry. To properly understand the interference effect, a quantitative model such as the Coulomb quantum-orbit strong field approximation [3, 17, 18] would be required.

## ORIENTATION AVERAGING

We will now examine the rotation of the target to account for different molecular orientations. The initial wave function is expanded into a sum over the spherical harmonics $\langle \mathbf{r} | \psi_0 \rangle = \sum_{\ell,m} a_{\ell,m} \psi_\ell^m (\mathbf{r})$ where $\psi_l^m (\mathbf{r}) \equiv R_{n,\ell} (r) Y_\ell^m (\theta_r, \phi_r)$ is each component, including a radial part and spherical harmonic. If we rotate it by the Euler

angles denoted $\varrho \equiv (\alpha, \beta, \gamma)$ [19], such that $|\psi_0(\varrho)\rangle \equiv \hat{R}(\varrho)\,|\psi_0\rangle$, where $\hat{R}(\varrho)$ is the rotation matrix, we get

$$\langle \mathbf{r} | \psi_0(\varrho) \rangle = \sum_{\ell, m, m'} a_{\ell,m} D^{(\ell)}_{m',m}(\varrho)\,\psi_\ell^{m'}(\mathbf{r}), \tag{11}$$

where $D^{(\ell)}_{m',m}(\rho)$ are elements of the Wigner rotation matrix.

To take into account both methods (QPROP and the SFA), we consider a general transition amplitude modified by the rotated initial state

$$M_f(\varrho) = \lim_{\substack{t \to \infty \\ t_0 \to -\infty}} \langle \psi_f | U(t,t_0) | \psi_0(\varrho) \rangle. \tag{12}$$

We are interested in the absolute value squared of the transition amplitude with the rotated initial state, averaged over orientations with a weighting factor $P(\varrho)$, i.e. $I[P, M_f] = \int d\varrho\, P(\varrho) |M_f(\varrho)|^2$, being $\int d\varrho \equiv \frac{1}{8\pi^2} \int_0^{2\pi} d\alpha \int_0^\pi d\beta \int_0^{2\pi} d\gamma \sin\beta$,

$$I[P, M_f] = \sum_{\ell, m, m', \lambda, \mu, \mu'} a_{\ell,m} a^*_{\lambda,\mu} \alpha^{\ell,\lambda}_{m,m',\mu,\mu'} M_{f;\ell,m'} M^*_{f;\lambda,\mu'}, \tag{13}$$

where $\alpha^{\ell,\lambda}_{m,m',\mu,\mu'} = \int d\varrho\, P(\varrho) D^{(\ell)}_{m',m} D^{(\lambda)*}_{\mu',\mu}$ and $M_{f;\ell,m} = \lim_{\substack{t \to \infty \\ t_0 \to -\infty}} \langle \psi_f | U(t,t_0) | \psi_\ell^m \rangle$.

For $P(\varrho) = 1$, i.e. for isotropic averaging, we have $\alpha^{\ell,\lambda}_{m,m',\mu,\mu'} = \frac{\delta_{m,\mu}\delta_{m',\mu'}\delta_{\ell,\lambda}}{2\ell+1}$ due to orthogonality of Wigner matrices and thus

$$I[1, M_f] = \sum_{\ell, m, m'} \frac{1}{2\ell+1} |a_{\ell,m}|^2 |M_{f;\ell,m'}|^2 \tag{14}$$

Note, this expression is symmetric with respect to inversion, so it is the same for opposite enantiomers, i.e. it is not enantio-sensitive due to the spherically symmetric Coulomb potential employed [20, 21].

Consider now $P(\varrho) = 3\cos^2\beta$ (the 3 is for normalization $\int d\varrho P(\varrho) = 1$), which corresponds to the orientation averaging of the photoelectron probability distribution for a partially aligned state with $\langle \cos^2(\beta) \rangle = 0.6$. In this case, since $D^{(2)}_{0,0}(\varrho) = (3\cos^2\beta - 1)/2$, we have $P(\varrho) = 2D^{(2)}_{0,0}(\varrho) + 1$ and consequently

$$\alpha^{\ell,\lambda}_{m,m',\mu,\mu'} = \delta_{m,\mu}\delta_{m',\mu'} \left[ 2(-1)^{m'-m} \begin{pmatrix} 2 & \ell & \lambda \\ 0 & m' & -m' \end{pmatrix} \begin{pmatrix} 2 & \ell & \lambda \\ 0 & m & -m \end{pmatrix} + \frac{\delta_{\ell,\lambda}}{2\ell+1} \right], \tag{15}$$

where the bracketed terms are the Wigner 3-j symbols, which can be directly related to Clebsch–Gordan coefficients.

If we assume we are measuring the OAM, with $\psi_f = \psi_{l_v}$, further simplifications can be made to the above expressions. Since OAM is conserved, $m' = \mu' = l_v$, and we get,

$$I[1, M_{l_v}] = \sum_{\ell, m} \frac{1}{2\ell+1} |a_{\ell,m}|^2 |M_{f;\ell,l_v}|^2 \tag{16}$$

for isotropic averaging, and

$$I[3\cos^2\beta, M_{l_v}] = \sum_{\ell, \lambda, m} a_{\ell,m} a^*_{\lambda,m} \left[ 2(-1)^{l_v-m} \begin{pmatrix} 2 & \ell & \lambda \\ 0 & l_v & -l_v \end{pmatrix} \begin{pmatrix} 2 & \ell & \lambda \\ 0 & m & -m \end{pmatrix} + \frac{\delta_{\ell,\lambda}}{2\ell+1} \right] M_{f;\ell,l_v} M^*_{f;\lambda,l_v}, \tag{17}$$

for orientation averaging corresponding to $P(\varrho) = 3\cos^2\beta$.

## EXPERIMENTAL IMPLEMENTATION

The measurement of the OAM of a photoelectron is at the forefront of experimental capability, and thus, at the moment, is neither trivial nor standard. However, a lot of activity in the field of electron optics has resulted in an increasing variety of methods (inspired by their analogues in optics) for the measurement of the OAM of an electron wave. These range from simply diffracting the electron wave through a knife edge [22], to mode conversion with astigmatic lenses [23], fork holograms [24, 25], Dammann vortex grating [26], and OAM sorting using conformal

mapping [27, 28], among others. The OAM sorter [28] may be retro fitted to existing electron microscopes and is intended for analysing OAM components of an electron beam after interaction with a sample.

This variety of methods, have built upon the electron optics found in transmission electron microscopes (TEMs) [29], and thus it is within this context that an experimental proof of principle of PEVD seems most feasible in the near future. Incorporating OAM measurements into strong field experiments (see Refs. [10, 30] for previous discussion on this) does require certain challenges to be met. For example, the photoelectrons must be properly aligned with the OAM analyser and accelerated to high energies for proper measurement. Both these considerations are completely resolved by the electron optics already present in a TEM. As such, we propose a method to test for PEVD that requires the least experimental development. To use chiral molecules as an electron source in a TEM already fitted with an OAM measurement stage. The chiral molecules would be photoionized by a laser pulse, then the electrons emitted in the direction of the electron condenser would be properly focused in order to be analysed in the OAM measurement stage. Note that ultrafast TEMs have already made use of short laser pulses to trigger the electron gun [31] and thus the experimental novelty of our proposal consists in the replacement of the usual electron source (e.g. a metal nanotip) by a chiral medium.

The data and plotting scripts for all figures used in this manuscript is freely available on a Zenodo database. The code written for this manuscript has been made open source and is available as a release on Zenodo and the repository is on GitHub.

---


\end{document}